# Band Alignments of Emerging Wurtzite BAlN and BGaN Semiconductors

Ahmad Al Sulami, Feras Alqatari, and Xiaohang Li

King Abdullah University of Science and Technology (KAUST), Advanced Semiconductor Laboratory, Thuwal 23955-6900, Saudi Arabia

**Abstract:**

The wurtzite III-Nitrides family of semiconductors, which include the compounds GaN, InN, and AlN, along with their derivative ternary alloys, is highly priced for its wide range of bandgaps, lattice constant tunability, high breakdown voltages, and thermal and chemical stability. The incorporation of wurtzite $B_xAl_{1-x}N$ and $B_xGa_{1-x}N$ ($0 \leq x \leq 1$) ternary alloys into this family introduces an even larger range of bandgaps, lattice constants, and refractive indices, which indicates their potential in the fields of optoelectronics and power devices. An important parameter in the design of cutting edge devices is the band alignment between the different alloys. In our work, the natural band offset values between wz-$B_xAl_{1-x}N$ and wz-$B_xGa_{1-x}N$ alloys were investigated using ab initio simulations. The Vienna Ab initio Simulation Package was used to perform density functional theory calculations in order to obtain lattice parameters, band gap energies, and relative electrostatic potential lineups. Through these calculations, we were able to quantify the natural band offset values for the materials of interest, and as such were able to identify some general qualitative features associated with the different alloys we studied. As the growth and fabrication of wz-BAlN and wz-BGaN crystals matures, we hope that our results can provide a theoretical basis for design and analysis of cutting-edge devices.

**Introduction:**

Semiconductors are the backbone of our modern society. Electronic devices that enabled efficient computing, wireless communications, power distribution, and efficient lighting owe their functionality and capabilities to the fact that they are based on semiconductor materials[1]. Within the compound III-V semiconductor family, the wurtzite III-Nitrides AlN, GaN, InN, and their derivative ternary compounds, possess unique properties and high flexibility, and as such they have generated significant interest starting from the late 1980s up until today. Among the various reasons wurtzite III-Nitride are valued are the bandgap tunability[2], high breakdown voltages[3], high thermal and chemical stability, and high carrier mobilities[4]. In the late 1980s and early 1990s, GaN resurfaced as a major semiconductor compound. Three Japanese researchers, Hiroshi Amano, Isamu Akasaki, and Shuji Nakamura, introduced major breakthroughs in epitaxial growth techniques and p- doping for GaN, and with that initiated the era of III-Nitride based optoelectronics[5,6]. The significance of these contributions is crowned by the awarding of the 2014 Nobel Prize in Physics to these three researchers.

Among the hot topics in the field of semiconductor devices nowadays is UV-emitters. UV light emitters in the form of LEDs and lasers are expected to provide utility in the fields of air purification, water and food sterilization, biochemical sensing, and photonics- based communications[4,7-10]. The AlGaN family of alloys provides a promising basis for such devices, and researchers are still investigating and optimizing device designs and growth methods to extract the best efficiency and performance from AlGaN materials.

In light of the recent trends in III-Nitride semiconductor research, the investigation of the element Boron, which has a chemical symbol B, as an alloying component seems to be a natural extension to previous ideas. In recent years, researchers have been experimenting with two new families of alloys, BGaN and BAlN[11-15]. Alloying BN with GaN or AlN gives us more options for lattice engineering and matching[15], which is a significant factor in the performance of semiconductor devices, and provides utility for a new range of bandgap energies, which provides more device design flexibility and perhaps new applications. This has the potential to improve AlGaN devices, in particular it can solve the problem of lattice mismatch and induced polarization fields currently present in such devices.

In this work, the natural band alignment between the alloys of the two systems BAlN and BGaN was investigated theoretically using first principle calculations. The structures for the alloys of interest were optimized numerically, and based on this optimized structure we were able to perform bulk simulations to determine the energy bandgap and average electrostatic potential for the respective alloys. Combining these results with surface calculations, we were able to infer the numerical values of the valence and conduction band offsets, which allowed for a proper categorization of the band alignment types for any pair of BAlN and BGaN alloys.

**Methodology and Computational Setup:**

The calculations in this study were performed using the Vienna ab initio Simulation Package VASP script as implemented in the MedeA interface. The computational setup for band alignment calculations was conceived after a careful review of multiple papers[16-24], which form the motivational basis for our work. Through the bulk simulations, we optimized the geometry, obtained the bandgap energies, and inferred the average electrostatic potential for a given alloy.

From the slab models, we have estimated the potential lineup between our different materials. The main equation we used to compute the VBO is:

$$\Delta E_V = E_v^{(1)} - E_v^{(2)} = \left(E_v - V_{average}\right)^{(1)} - \left(E_v - V_{average}\right)^{(2)} + \Delta V \quad (1)$$

With (i) denoting a property of material i, $E_v$ the valence band edge energy, $V_{average}$ the average electrostatic field in the bulk material, and $\Delta V$ the difference in the aligned electrostatic potentials. The CBO is relatively simpler to obtain:

$$\Delta E_c = E_c^{(1)} - E_c^{(2)} = E_g^{(1)} - E_g^{(2)} - \Delta E_v \quad (2)$$

Where $E_g^{(i)}$ is the bandgap energy for material (i). Here we adhere to the convention that a positive $\Delta E_c$ indicates material 2 has a higher CBM compared to material 1.

The wurtzite bulk materials were constructed as follows:

- Binary alloys were constructed in primitive four atom cells

- Ternary alloys with equal cation compositions, e.g. $B_{0.5}Al_{0.5}N$, were constructed in chalcopyrite-like cells composed of sixteen atoms. This structure is characterized by the fact that the anions are surrounded by two atoms of each cation

- Ternary alloys with compositions of 25% and 75% were constructed using luzonite-like cells composed of sixteen atoms. In this structure, each anion is surrounded by one atom of the higher concentration cation species and three atoms of the other cation species. Intermediate compositions 12.5%, 37.5%, 62.5%, and 87.5% are given by the same luzonite-like structure, with one of the ratio of 3:1 of 2:2 of cations around a given anion, but with a 4:0 ratio of cations in the adjacent anion

These choices are guided by the previous work of our colleagues, as it is discussed in their publication[14].

Throughout our simulations, we have kept a few choices invariant. We have used plane augmented waves pseudopotentials to describe atomic cores, with the inclusion of d-orbital electrons as core electrons in Ga. Structural optimization calculations were performed by relaxing the atomic positions to Hellmann-Feynman forces below a threshold of $2 \times 10^{-2} \frac{eV}{\text{Å}}$. All energy calculations were set for a convergence criterion of $10^{-5} \frac{eV}{atom}$. Because all of the alloys we are interested in possess a hexagonal symmetry, the k-meshes used for reciprocal space sampling were chosen to be $\Gamma-$ centerd Monkhorst-Pack grids, with the size chosen as appropriate to the cell choice. For all calculations, spin-orbital coupling and magnetic effects were ignored, since these are negligible for the materials of interest.

To ensure our simulations are well-behaved numerically, we have performed a series of convergence tests on a set of parameters on which our simulations depend. Specifically, we have performed convergence analysis on the k-mesh size, planewave cutoff energy, number of repeating units in a material slab, and the vacuum width in the slab model. Based on these preliminary simulations, we have decided to set the k-mesh at $6 \times 6 \times 6$ for ternary alloys and $8 \times 8 \times 6$ for binary alloys. A planewave cutoff energy of 520 eV was found to produce sufficiently accurate results. For the slab calculations, a slab thickness of 13 atomic layers and a vacuum thickness of 7 atomic layers was chosen. The choice for slab thickness is consistent with the finding of Tsai et al [25].

**Results and Discussion:**

Structural optimization was performed using the GGA exchange-correlation functional, specifically the Perdew-Burke-Ernzerhof version revised for solids, which is known as GGA-PBEsol. The use of such functional is highly successful for solids, and is more accurate and reliable than the LDA functional[23]. In table 1 we list the lattice constants we obtained from our setup, along with experimental values, for the binary III-Nitrides.

Table 1. Structural parameters for wurtzite GaN, AlN and BN.

|     |           | $a$ (Å) | $c$ (Å) |
|-----|-----------|---------|---------|
| GaN | This work | 3.18    | 5.17    |
|     | Ref [24]  | 3.18    | 5.166   |
| AlN | This work | 3.11    | 4.98    |
|     | Ref [24]  | 3.11    | 4.93    |
| BN  | This work | 2.56    | 4.23    |
|     | Ref [24]  | 2.55    | 4.21    |

In Figure 1 we plot our results for the lattice parameters of all the alloys, along with a second order polynomial fit to investigate Vegard's law.

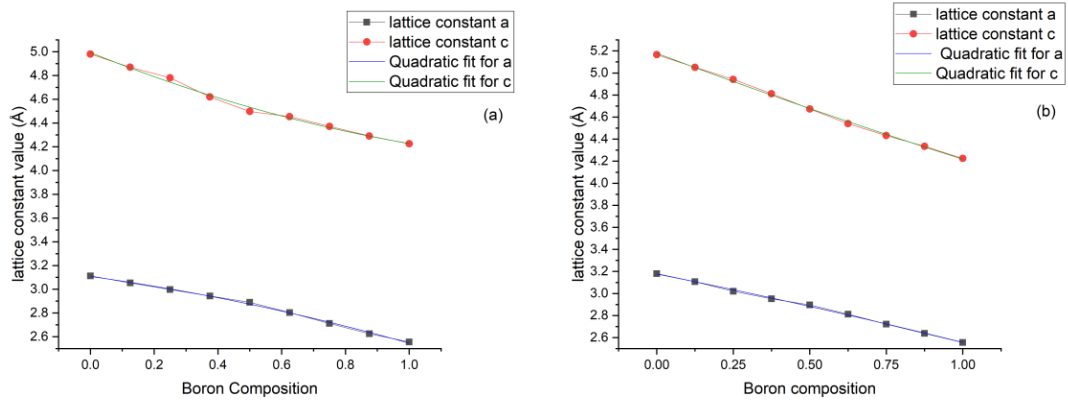

Fig. 1. Computational results for the lattice constants as a function of the B composition for wurtzite (a) BAlN and (b) BGaN. The quadratic polynomial fits are included in the plots. The polynomial fit gave the following results and coefficients of determination:

$$a_{B_xAl_{1-x}N}(x) = -0.19x^2 - 0.37147x + 3.10783 \text{ Å}, \quad r^2 = 0.99814$$

$$c_{B_xAl_{1-x}N}(x) = 0.30944x^2 - 1.07508x + 4.99353 \text{ Å}, \quad r^2 = 0.99412$$

$$a_{B_xGa_{1-x}N}(x) = -0.06637x^2 - 0.55397x + 3.17695 \text{ Å}, \quad r^2 = 0.99881$$

$$c_{B_xGa_{1-x}N}(x) = 0.08171x^2 - 1.04191x + 5.17853 \text{ Å}, \quad r^2 = 0.99858$$

These number are in good agreement with the results in the reference[8], and the high coefficients of determination indicate that our simulated material structures follow Vegard's law very closely.

The optimized structures were used to perform bulk and slab calculations. It is a known fact that LDA and GGA functionals underestimate the bandgap energies of semiconductors, especially wide-bandgap semiconductors including a large alloy range of the wurtzite III-Nitrides[26]. As such we chose to perform the bulk calculations used the Heyd, Scuseria, and Ernzerhof hybrid functional with a 0.25 mixing parameter, commonly referred to as HSE06 .

We performed slab calculations in order to establish a reference for the bulk results. we choose to build our slab model as an interface between 6 bulk cell repetitions of the material, which corresponds to 13 atomic layers, interfaced with 4 bulk cell repetitions of vacuum, corresponding to 7 atomic layers in width. It is worth noting that these convergence results are in agreement with the findings of Moses et al[23], in particular the result that 12 atomic layers produce relative potential positions with respect to vacuum that are no more than 0.03 eV different from the positions obtained with a 28 atomic layer thick slab. Figure 2 shows a demonstration of our setup for potential lineup calculation.

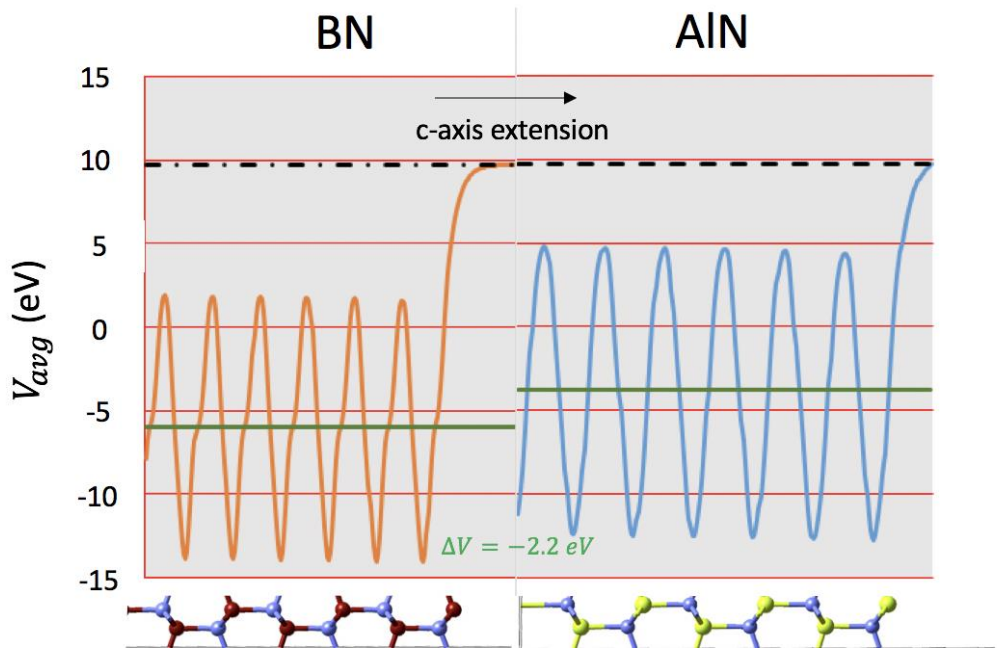

Fig. 2. A demonstration of the potential line-up setup for the example interface BN/AlN

To find the relative alignment of the average electrostatic potential between two materials, we combine the two respective slab calculations by aligning the vacuum levels to establish an appropriate absolute reference. The choice of a numerical number for the vacuum is arbitrary and will not affect the relative alignment. As our goal in this study is to calculate the natural unstrained band alignment, the choice of measuring the slab potentials relative to vacuum independently is justified. This justificationcan be reviewed in more detail in the paper published by Su-Huai Wei et al[24]. In this paper, the researchers highlight the validity of such approximation by demonstrating the transitivity law, which states that the valence band offset between two materials AX and CX can be expressed as follows:

$$\Delta E_v(AX \backslash CX) = \Delta E_v(AX \backslash BX) + \Delta E_v(BX \backslash CX)$$

Where BX is a third material. According to the results and discussions in the paper, this hold very well for natural band alignment calculations, and indicates that the potential lineup can be obtained with high accuracy without establishing a direct contact between the two materials.

For the potential lineup calculations, k-meshes were chosen to be of size $8 \times 8 \times 2$ for the binary alloys and $6 \times 6 \times 2$ for the ternary alloys, and the exchange-correlation functional of choice was GGA-PBE. Table 2 summarizes the results obtained from bulk and slab calculations.

Table 2. Band gap energy, optical transition, average electrostatic potential, and relative potential position for all materials.

| Alloy | Bandgap (eV) | Optical transition | $VBM-V_{avg}$ (eV) | $V_{vac}-V_{avg}$ (eV) |
|---|---|---|---|---|
| **AlN** | 5.56 | direct | 15.92 | 13.17 |
| **B$_{0.125}$Al$_{0.875}$N** | 5.18 | direct | 16.165 | 13.25 |
| **B$_{0.25}$Al$_{0.75}$N** | 5.56 | direct | 16.23 | 13.53 |
| **B$_{0.375}$Al$_{0.625}$N** | 4.971 | indirect | 16.31 | 13.64 |
| **B$_{0.5}$Al$_{0.5}$N** | 5.42 | indirect | 16.96 | 13.80 |
| **B$_{0.625}$Al$_{0.375}$N** | 5.74 | indirect | 17.21 | 13.95 |
| **B$_{0.75}$Al$_{0.255}$N** | 6.49 | indirect | 17.75 | 14.45 |
| **B$_{0.875}$Al$_{0.125}$N** | 6.52 | indirect | 18.29 | 14.62 |
| **BN** | 6.56 | indirect | 18.03 | 15.39 |
| **B$_{0.875}$Ga$_{0.125}$N** | 6.59 | indirect | 16.80 | 14.355 |
| **B$_{0.75}$Ga$_{0.255}$N** | 5.38 | indirect | 16.24 | 13.815 |
| **B$_{0.625}$Ga$_{0.375}$N** | 5.27 | indirect | 16.02 | 13.185 |
| **B$_{0.5}$Ga$_{0.5}$N** | 5.00 | indirect | 16.05 | 12.1675 |
| **B$_{0.375}$Ga$_{0.625}$N** | 4.36 | direct | 15.83 | 10.735 |
| **B$_{0.25}$Ga$_{0.75}$N** | 3.92 | direct | 15.77 | 10.28 |
| **B$_{0.125}$Ga$_{0.875}$N** | 3.54 | direct | 15.44 | 9.95 |
| **GaN** | 3.30 | direct | 15.27 | 11.59 |

Combining the computed parameters through equations (1) and (2), we can obtain the relative VBO and CBO values for the whole range of alloys. The results are summarized in table 3 and figure 3 below. Note that, in both result forms, we have chosen to set the valence band maximum of AlN at 0 eV, meaning that all the band edge values are measured with respect to the valence band maximum of AlN.

Table 3. Valence band edge values for BAlN and BGaN values

| Alloy | Valence band edge (eV) | Conduction band edge (eV) |
|---|---|---|
| AlN | 0 | 5.56 |
| $B_{0.125}Al_{0.875}N$ | 0.165 | 5.35 |
| $B_{0.25}Al_{0.75}N$ | -0.055 | 5.51 |
| $B_{0.375}Al_{0.625}N$ | -0.08 | 4.89 |
| $B_{0.5}Al_{0.5}N$ | 0.41 | 5.83 |
| $B_{0.625}Al_{0.375}N$ | 0.507 | 6.25 |
| $B_{0.75}Al_{0.255}N$ | 0.545 | 7.035 |
| $B_{0.875}Al_{0.125}N$ | 0.915 | 7.435 |
| BN | -0.11 | 6.45 |
| $B_{0.875}Ga_{0.125}N$ | -0.31 | 6.28 |
| $B_{0.75}Ga_{0.255}N$ | -0.333 | 5.05 |
| $B_{0.625}Ga_{0.375}N$ | 0.08 | 5.35 |
| $B_{0.5}Ga_{0.5}N$ | 1.128 | 6.13 |
| $B_{0.375}Ga_{0.625}N$ | 2.33 | 6.69 |
| $B_{0.25}Ga_{0.75}N$ | 2.74 | 6.66 |
| $B_{0.125}Ga_{0.875}N$ | 2.735 | 6.28 |
| GaN | 0.925 | 4.23 |

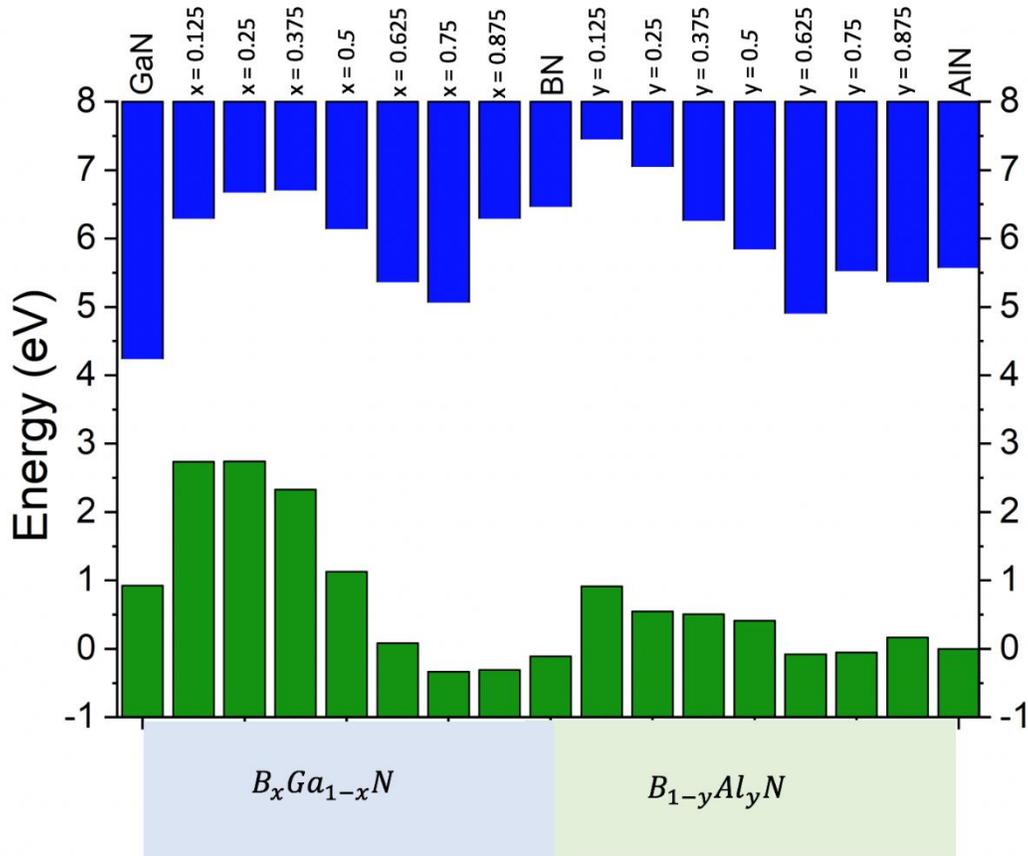

Fig. 3. A summary of the band alignment results. Between each bar is a $\frac{1}{8}$ increment change in compositions. The valence band of AlN has been set at the 0 eV for convenience.

One of the interesting observations we can make is that for BGaN alloys, the dominant band alignment is type-II. Another interesting observation is than the two alloy systems behave differently; for BAlN, the valence band offset doesn't take large values in single increments, while for BGaN alloys, large numerical values for the valence band offset occur.

**Conclusions:**

In summary, we have studied the natural band alignment between alloys in the emerging III-Nitride systems BAlN and BGaN. We have utilized DFT simulations as implemented in the VASP script, and have used hybrid functionals in order to accurately report the bulk properties of these alloys and their parent binaries. We have performed a set of preliminary tests that enabled us to optimize for efficiency and accuracy simultaneously, in particular in our choice of k-point densities and potential line-up settings. We have reported structural parameters for BAlN and BGaN alloys, and we have seen how they followed Vegard's law to a high degree, as is the case with other compound semiconductors. We have used the standard potential line-up method to calculate the band offsets, but our approach differed from previous reports in its use of slab models that are extended in the polar c-direction. Our computational results make some interesting predictions, including the dominance of type-II alignment in GaN/BGaN interfaces for a large range of B compositions.

The KAUST authors would like to acknowledge the support of like to acknowledge the support of KAUST Baseline Fund BAS/1/1664-01-01, GCC Research Council Grant REP/1/3189-01-01, and Competitive Research Grants URF/1/3437-01-01 and URF/1/3771-01-01.